\documentclass[preprint,aps,prd,showpacs,nofootinbib]{revtex4}
\usepackage{mathrsfs}
\usepackage{amsmath}
\usepackage{graphicx}
\usepackage{caption}
\usepackage{subfigure}
\usepackage{bm}
\usepackage{epstopdf}
\usepackage{float}
\usepackage{latexsym}

\newcommand{\bea}{\begin{eqnarray}}
\newcommand{\eea}{\end{eqnarray}}
\newcommand{\beq}{\begin{equation}}
\newcommand{\eeq}{\end{equation}}
\newcommand{\nn}{\nonumber}

\def\/{\over}

\begin{document}
\title{\bf Spontaneous excitation of a static atom in a thermal bath in cosmic string spacetime}
\author{Huabing Cai$^{1}$, Hongwei Yu$^{1,2}$ and Wenting Zhou$^{1}$ }
\affiliation{$^{1}$Center for Nonlinear Science and Department of Physics, Ningbo
University, Ningbo, Zhejiang 315211, China\\
$^{2}$Synergetic Innovation Center for Quantum Effects and Applications, Hunan Normal University, Changsha, Hunan 410081, China}

\begin{abstract}
We study the average rate of change of energy for a static atom immersed in a thermal bath of electromagnetic radiation
in the cosmic string spacetime and  separately calculate the contributions of thermal fluctuations and radiation reaction. We find that the transition rates are  crucially dependent on  the atom-string distance and
 polarization of the atom and they in general oscillate as the atom-string distance varies. Moreover, the atomic transition rates in the cosmic string spacetime can be larger or smaller than those in Minkowski spacetime contingent upon the atomic polarization and position. In particular, when located on the string, ground-state atoms can make a transition to excited states only if they are polarizable parallel
to the string, whereas  ground state atoms  polarizable only perpendicular to the string are stable
as if they were in a vacuum, even if  they are immersed in a thermal bath. Our results suggest that the influence of a cosmic
string is very similar to that of a reflecting boundary in Minkowski spacetime.
 \end{abstract}

\maketitle

\section{INTRODUCTION}

Spontaneous emission is one of the most important phenomena in the interaction of atoms with radiation, and it can be attributed to vacuum fluctuations~\cite{Welton48,Compagno83}, or radiation reaction~\cite{Ackerhalt73}, or a combination of them~\cite{Milonni88,Milonni75}.
So far, a lot of  efforts have been made to resolve the ambiguity in the underlying  mechanism regarding the radiative properties of atoms~\cite{Milonni75,Vleck24,Dirac27,Ackerhalt73,Senitzky73,Milonni73,Ackerhalt74,Milonni752}. In this regard,  Dalibard, Dupont-Roc
and Cohen-Tannoudji (DDC) suggested a resolution which distinctively separates the contributions of vacuum fluctuations and radiation reaction by choosing a symmetric ordering
between the operators of the dynamical variables of the atom and the field which ensures the Hermitianity of the Hamiltonians of
vacuum fluctuations and radiation reaction~\cite{Dalibard8284}. Later, the DDC formalism was generalized  to investigate the
radiative properties of atoms in different circumstances, such as a non-inertial atom in interaction
with various quantum fields~\cite{Audretsch94,Audretsch951,Passante98,Yu-na05,Yu-na061,Yu-na062,Yu-na07,Rizzuto07,Rizzuto09,Yu-na10,Rizzuto11,Yu-na12},
and an inertial atom immersed in a thermal bath~\cite{Tomazelli03,Yu-th0912,Yu-th10}. %Studies on these topics predicts the existence of an important physical phenomenon-spontaneous excitation.
In both cases, as the contribution of the fluctuations of the quantum field and that of the radiation
reaction to the rate of change of the atomic energy no longer cancel  completely, an atom  in the ground state can make a
spontaneous transition to excited states.

In recent years, investigations on the radiative properties of atoms have been extended to curved spacetime~\cite{Iliadakis,Yu-cur07,Yu-cur08,Yu-cur12}. It is interesting to note that these studies along with those for non-inertial atoms in flat spacetime have shed light on the nature of the Hawking radiation of black holes, the Gibbons-Hawking effect of de Sitter space as well as the Unruh effect related to uniformly accelerated observers as atoms can serve as a model of realistic particle detectors.
%Our interest in this topic lies in that atoms may be considered as a detector to sense the properties of curved spacetime.
In this paper, we plan to study the spontaneous excitation of static atoms in yet another typical curved spacetime, i.e., the spacetime of a  cosmic string.  In comparison to other spacetimes, the cosmic string spacetime is characterized by its  structure with non-trivial topology, a planar deficit angle, to be specific. Although now  much remains to be done to fully understand the behavior of strings, people are convinced that they may raise a number of issues in fundamental physics, for example,  gravitational effects
such as lensing of distant objects and conical bremsstrahlung~\cite{Vilenkin81,Aliev89,Aliev93}. Interestingly, one can also use atoms to sense a cosmic string. In this respect,  J. Audretsch, et al.,
studied the spontaneous emission and the Lamb shift of an atom in a toy model where the atom is assumed to be coupled to vacuum quantum scalar fields in the cosmic string spacetime and found that the spontaneous emission rate
is modified by the presence of a cosmic string~\cite{Audretsch952}.
%Besides, as the modes of quantized fields propagating in the curved spacetime are restricted by the non-trivial topological structures, boundary effects are expected to appear.
Recently, a number of authors have studied the Casimir effect and Casimir-Polder force
% acting on a polarizable microparticle
  in a  more realistic situation where the atom interacts with electromagnetic vacuum fluctuations in the geometry of a straight
cosmic string% and found some peculiar properties
~\cite{Saharian11,Saharian121,Saharian122}.  In this paper, we plan to study the spontaneous excitation and emission of
 a static atom immersed in a thermal bath of electromagnetic radiation in the vicinity of a straight cosmic string, where the atom is coupled to quantum electromagnetic fields rather than scalar fields in~\cite{Audretsch952}.

The paper is organized as follows. In section \uppercase\expandafter{\romannumeral2}, we introduce the quantization of electromagnetic fields in cosmic
string space-time. In section \uppercase\expandafter{\romannumeral3}, we generalize the DDC formalism to study the average rate of change of the atomic
energy in the cosmic string spacetime. In section \uppercase\expandafter{\romannumeral4}, we concretely calculate the average rate of change of a static
atom immersed in a thermal bath in the cosmic string spacetime and discuss how the conical deficit angle affects the rate of change of atomic energy. Finally in
section \uppercase\expandafter{\romannumeral5}, we give some concluding remarks. Throughout the paper, we adopt the natural unit, $\hbar=c=1$, and
let the Boltzmann constant $k_B=1$.

\section{Quantum electromagnetic field in the cosmic string space-time}

The metric of a static, straight cosmic string lying along the $z$-direction in the cylindrical coordinate system is given by
\beq
d s^2=d t^2-d\rho^2-\rho^2d\theta^2-d z^2
\eeq
where $0\leq\theta<\frac{2\pi}{\nu}$, $\nu=(1-4G\mu)^{-1}$ with $G$ and $\mu$ being the Newton's constant and the mass per unit
length of the string respectively.  The Lagrangian density of the electromagnetic field can be written as
\beq
\mathcal{L}=\sqrt{-g}\biggl[-\frac{1}{4}F^{^{\mu\nu}}F_{_{\mu\nu}}-\frac{1}{2}(A^{\mu}_{\;;\mu})^2\biggr]\;.
\eeq
The quantization  of the field is to be carried out in the Feynman gauge %under which the gauge condition is
\beq
A^{\mu}_{\;;\mu}=0\;.
\eeq
Inserting the above Lagrangian density into the Euler-Lagrangian equation, we obtain %the equation for the electromagnetic field
\beq
F^{\mu\nu}_{\;\;\;\;;\nu}=0\;.
\eeq
In terms of the vector potential of the electromagnetic field, the above equation becomes
\beq
\Box A_{\rho}-\frac{2}{\rho^{3}}\partial_{\theta}A_{\theta}-\frac{1}{\rho^{2}}A_{\rho}=0\;,\label{equation for A rho}
\eeq
\beq
\Box A_{\theta}-\frac{2}{\rho}\partial_{\rho}A_{\theta}+\frac{2}{\rho}\partial_{\theta}A_{\rho}=0\;,\label{equation for A theta}
\eeq
\beq
A_{z}=\Box A_{t}=0\label{equation for A_z and A_t}
\eeq
with
\beq
\Box =\Delta-\partial_{tt}^{2}, \quad \Delta=\frac{1}{\rho}\partial_{\rho}(\rho\partial_{\rho})
+\frac{1}{\rho^{2}}\partial_{\theta\theta}^{2}+\partial_{zz}^{2}\;.
\eeq
To solve
Eqs.~(\ref{equation for A rho})-(\ref{equation for A_z and A_t}), we firstly decouple the field equations by introducing
the spin-weighted components of the vector potential~\cite{Aliev89}, i.e., define
\beq
A_{\xi}=\frac{1}{\sqrt{2}}(A_{\rho}+\frac{i\xi}{\rho}A_{\theta})\quad {\text {for}}  \quad \xi=\pm 1,\label{relation 12}
\eeq
\beq
A_{\xi}=A_{z},  A_{t} \quad\quad\;  {\text{for}} \quad \xi=3, 0\label{relation 34}\;.
\eeq
Then the decoupled field equations
can be collectively written as
\beq
\Box_{\xi}A_{\xi}=0
\eeq
with
\beq
\Box_{\xi}=\Delta_{\xi}-\partial_{tt}^{2}\;,
\eeq
\beq
\Delta_{\xi}=\frac{1}{\rho}\partial_{\rho}(\rho\partial_{\rho})-\frac{1}{\rho^{2}}L_{3}^{2}+\partial_{tt}^{2}\;,
\eeq
\beq
L_{3}=-i\partial_{\theta}+\xi\;.
\eeq
The normal modes for the independent components, $A_{\xi}$, are
\beq
f_{\xi j}(x)=f_{\xi j}(\vec{x})e^{-i\omega t}\label{normal modes}
\eeq
with
\beq
f_{\xi j}(\vec{x})=\frac{1}{2\pi}\sqrt{\frac{\nu}{2\omega}}J_{|\nu m+\xi|}(k_{\bot}\rho)
e^{i(\nu m\theta+k_{3}z)}\;,\label{normal modes spacial component}
\eeq
where the symbol $J$ denotes BesselJ function, the subscript $j=(k_{3},k_{\bot},m)$ and $\omega=\sqrt{k_{3}^{2}+k_{\bot}^{2}}$. The modes are normalized according to
\beq
\int d^{3}\bm{x} f_{\xi j}^{\ast}(x)(i\overleftrightarrow{\partial_{t}})f_{\xi j'}(x)=\delta_{j,j'}
=\delta_{m,m'}\delta(k_{3}-k_{3}')\frac{\delta(k_{\bot}-k_{\bot}')}{\sqrt{k_{\bot}k_{\bot}'}}.
\eeq

In order to quantize the electromagnetic field, we define the canonically conjugate field $\Pi^{\mu}$ corresponding
to $A_{\mu}$ as
\begin{eqnarray}
\Pi_{\mu}=\frac{1}{\sqrt{-g}}\frac{\partial{\mathcal{L'}}}{\partial{A^{\mu}_{\;\;;0}}}=-A_{\mu}^{\;\;;0
}
\end{eqnarray}
in which $\mathcal{L}'$ describes the dynamics of the electromagnetic field and it is obtained by discarding
a four-divergence term in $\mathcal{L}$ which has no influence on the field equations.  We impose the
following equal-time commutation relations for the field operator $A^{\mu}$ and $\Pi^{\mu}$
\beq
[A_{\mu}(t,\vec{x}),A_{\nu}(t,\vec{x})]=[\Pi_{\mu}(t,\vec{x}),\Pi_{\nu}(t,\vec{x})]=0,\label{relation between AA}\\
\eeq
\beq
[A_{\mu}(t,\vec{x}),\Pi^{\nu}(t,\vec{x}')]=i\delta_{\mu}^{\nu}\delta^{3}(\vec{x}-\vec{x}')\;.\label{relation between Api}
\eeq
Now we expand the field operator in terms of the complete set of normal modes (see Eq.~(\ref{normal modes})),
\beq
A_{\xi}(t,\vec{x})=\int d \mu_j\;[\;c_{\xi j}(t)f_{\xi j}(\vec{x})+c_{-\xi j}^{\dag}(t)f_{-\xi j}^{\ast}(\vec{x})]\label{Axi}
\eeq
in which
\beq
\int d \mu_j=\sum _{m=-\infty}^{\infty}\int_{-\infty}^{\infty}d k_{3}\int_{0}^{\infty}d k_{\perp}k_{\perp}\;,
\eeq
and $c_{\xi j}(t)=c_{\xi j}(0)e^{-i\omega t}$ and $c^{\dag}_{-\xi j}=c^{\dag}_{-\xi j}(0)e^{i\omega t}$ are respectively the
annihilation and creation operators for a photon with quantum numbers $(k_{3},k_{\perp},m)$ at time $t$. One can show that
\bea
c_{\xi j}(0)&=&i\int d^3\vec{x} f^{\ast}_{\xi j}(t,x)\overleftrightarrow{\partial_{t}}A_{\xi}(t,x)\;,\\
c^{\dag}_{\xi j}(0)&=&-i\int d^3\vec{x} f_{\xi j}(t,x)\overleftrightarrow{\partial_{t}}A_{\xi}(t,x)\;.
\eea
Now by using the relations Eqs.~(\ref{relation 12}), (\ref{relation 34}), (\ref{relation between AA}), (\ref{relation between Api}),
the commutation relations of the annihilation and creation operators are found to be
\bea
&&[c_{\xi j}(t),c_{\xi j'}^{\dag}(t)]=\delta_{j,j'}\quad\; for \quad\xi=\pm 1,3,\\
&&[c_{0 j}(t),c_{0 j'}^{\dag}(t)]=-\delta_{j,j'} \;\;
 for \quad \xi=0.
\eea
Here let us point out that  a minus sign in the commutation relations for $\xi=0$  in Eq.~(26), which is missing  in Ref.~\cite{Skarzhinsky94}, has been
added.

Finally by calculating the $T_{00}$ component of the stress tensor of the quantum electromagnetic field, we obtain
the Hamiltonian operator of the field
\beq
H_{F}=\int d \mu_j\;\omega_{j}(c_{+j}^{\dag}c_{+j}+c_{-j}^{\dag}c_{-j}+c_{3j}^{\dag}c_{3j}-c_{0j}^{\dag}c_{0j})\;.
\eeq

\section{The generalized DDC formalism}

We consider a multi-level atom in interaction with the quantum electromagnetic field in a thermal bath in the cosmic string space-time. The Hamiltonian that
governs the evolution of the atom with respect to the proper time, $\tau$, is given by
\beq
H_{A}(\tau)=\sum_{n}\omega_{n}\sigma_{nn}(\tau)\;,\label{Hamiltonian of the atom}
\eeq
in which $\sigma_{nn}=|n\rangle \langle n|$ and $|n\rangle$ denotes a complete set of atomic stationary state with energy $\omega_n$.
The Hamiltonian of the quantum electromagnetic field in the proper time, $\tau$, is
\beq
H_{F}(\tau)=\int d \mu_j\;\omega_{j}(c_{+j}^{\dag}c_{+j}+c_{-j}^{\dag}c_{-j}+c_{3j}^{\dag}c_{3j}-c_{0j}^{\dag}c_{0j})\frac{dt}{d\tau}\;.
\label{Hamiltonian of the field}
\eeq
We assume that the atom interacts with the quantum electromagnetic field in the multipolar coupling scheme~\cite{Passante98}, so
the interaction Hamitonian can be written as
\beq
H_{I}(\tau)=-e\textbf{r}(\tau)\cdot\textbf{E}(x(\tau))=-e\sum_{mn}\textbf{r}_{mn}\cdot\textbf{E}(x(\tau))\sigma_{mn}(\tau)
\eeq
where $e$ is the electron electric charge, $e\textbf{r}$ is the atomic dipole moment, and $x(\tau)\leftrightarrow(t(\tau),\vec{x}(\tau))$
is the space-time coordinate of the atom in the cosmic string spacetime. The Hamiltonain that determines the time evolution of
the system (atom+field) is composed by the above three parts
\beq
H(\tau)=H_{A}(\tau)+H_{F}(\tau)+H_{I}(\tau).
\eeq

Starting from the above Hamiltonian, we can write out the Heisenberg equations for the dynamical variables of the atom and
the field. In the formal solutions, we can separate each solution of either the variable of the atom or the field into the
``free'' part which exists even in the vacuum, and the ``source" part which is induced by the interaction between the atom
and the field,
\bea
\sigma_{mn}(\tau)&=&\sigma_{mn}^{f}(\tau)+\sigma_{mn}^{s}(\tau)\;,\\
c_{\xi j}(t(\tau))&=&c_{\xi j}^{f}(t(\tau))+c_{\xi j}^{s}(t(\tau))\;,
\eea
where
\bea
\left\{
  \begin{array}{ll}
    c_{\xi j}^{f}(t(\tau))=c_{\xi j}(t(\tau_0))e^{-i\omega_{j}[t(\tau)-t(\tau_0)]}\;, \\
    c_{\xi j}^{s}(t(\tau))=-ie\int_{\tau_{0}}^{\tau}d\tau'\;[\textbf{r}(\tau')\cdot\textbf{E}(x(\tau')),c_{\xi j}^{f}(t(\tau))]\;,
  \end{array}
  \right.
\eea
and
\bea
\left\{
  \begin{array}{ll}
    \sigma_{mn}^{f}(\tau)=\sigma_{mn}^{f}(\tau_{0})e^{i\omega_{mn}(\tau-\tau_0)}\;, \\
    \sigma_{mn}^{s}(\tau)=-ie\int_{\tau_{0}}^{\tau}d\tau'\;[\textbf{r}(\tau')\cdot\textbf{E}(x(\tau')),\sigma_{mn}^{f}(\tau)]\;.
  \end{array}
\right.
\eea
Consequently, the free part and source part of the vector potential operator can be expressed as
\bea
A^f_{\xi}(t,\vec{x})&=&\int d \mu_j\;[\;c^f_{\xi j}(t) f_{\xi j}(\vec{x})+{c^{f\dag}_{-\xi j}}(t)f_{-\xi j}^{\ast}(\vec{x})]\;,\\
A_{\xi}^{s}(t,\vec{x})&=&-ie\int_{\tau_{0}}^{\tau}d\tau'\;[\textbf{r}^{f}(\tau')\cdot\textbf{E}^{f}(x(\tau')),A_{\xi}^{f}(x(\tau))]\;.
\eea
Notice that in the source parts of the above solutions, all operators on the right-hand side have been replaced
by their free parts, which are correct to the first order in $e$.

Taking the observable to be the energy of the atom, we obtain%The Heisenberg equation of the atomic Hamiltonian is
\beq
\frac{d H_A(\tau)}{d\tau}=-ie[\textbf{r}(\tau)\cdot\textbf{E}(x(\tau)),H_A(\tau)]\;.
\eeq
Now following  DDC~\cite{Dalibard8284}, we separate the field operator into the free part and
the source part, $\textbf{E}(x(\tau))=\textbf{E}^f(x(\tau))+\textbf{E}^s(x(\tau))$, and choose a symmetric ordering
between the operators of the variables of the atom and the field. Then we can identify the contributions of the free part and
the source part, i.e., the contributions of thermal fluctuations and radiation reaction,
\beq
\frac{d H_A(\tau)}{d\tau}=\biggl(\frac{dH_A(\tau)}{d\tau}\biggr)_{tf}+\biggl(\frac{dH_A(\tau)}{d\tau}\biggr)_{rr}
\eeq
with
\bea
\biggl(\frac{d H_A(\tau)}{d\tau}\biggr)_{tf}&=&-\frac{ie}{2}(\textbf{E}^{f}(x(\tau))\cdot[\textbf{r}^f(\tau),H_A(\tau)]
+[\textbf{r}^f(\tau),H_A(\tau)]\cdot\textbf{E}^{f}(x(\tau)))\;,\\
\biggl(\frac{d H_A(\tau)}{d\tau}\biggr)_{rr}&=&-\frac{ie}{2}(\textbf{E}^{s}(x(\tau))\cdot[\textbf{r}^f(\tau),H_A(\tau)]
+[\textbf{r}^f(\tau),H_A(\tau)]\cdot\textbf{E}^{s}(x(\tau)))\;.
\eea
Averaging the above two equations over the state of the field, $|\beta\rangle$, and the atomic state, $|a\rangle$,
 we obtain, after some simplifications, the contributions of thermal fluctuations and radiation reaction to the average
rate of change of the atomic energy,
\bea
\biggl\langle \frac{d H_{A}(\tau)}{d\tau}\biggr\rangle_{tf}&=&
2ie^{2}\int_{\tau_{0}}^{\tau}d\tau'C_{ij}^{F\beta}(x(\tau),x(\tau'))\frac{d}{d\tau}\chi^{ijA}_{b}(\tau,\tau')\;,
\label{formula for tf contribution}\\
\biggl\langle \frac{d H_{A}(\tau)}{d\tau}\biggr\rangle_{rr}&=&
2ie^{2}\int_{\tau_{0}}^{\tau}d\tau'\chi_{ij}^{F\beta}(x(\tau),x(\tau'))\frac{d}{d\tau}C^{ijA}_{b}(\tau,\tau')\;,
\label{formula for rr contribution}
\eea
where $C_{ij}^{F\beta}(x(\tau),x(\tau'))$ and $\chi_{ij}^{F\beta}(x(\tau),x(\tau'))$ are respectively the symmetric
correlation function and the linear susceptibility function of the quantum electromagnetic field defined as
\bea
C_{ij}^{F\beta}(x(\tau),x(\tau'))&=&\frac{1}{2}\langle\beta|\{E_{i}^{f}(x(\tau)),E_{j}^{f}(x(\tau'))\}|\beta\rangle\;,\label{definition for CF}\\
\chi_{ij}^{F\beta}(x(\tau),x(\tau'))&=&\frac{1}{2}\langle\beta|[E_{i}^{f}(x(\tau)),E_{j}^{f}(x(\tau'))]|\beta\rangle\;,\label{definition for ChiF}
\eea
and $C^{ijA}_{b}(\tau,\tau')$ and $\chi^{ijA}_{b}(\tau,\tau')$ are the two statistical functions of the atom
in state $|b\rangle$  which are defined as follows
\bea
C^{ijA}_{b}(\tau,\tau')&=&\frac{1}{2}\sum_{d}[\langle b|r^{i}(0)|d\rangle\langle d|r^{j}(0)|b\rangle e^{i\omega_{bd}(\tau-\tau')}
+\langle b|r^{j}(0)|d\rangle\langle d|r^{i}(0)|b\rangle e^{-i\omega_{bd}(\tau-\tau')}]\;,\nn\\\label{cA} \\
\chi^{ijA}_{b}(\tau,\tau')&=&\frac{1}{2}\sum_{d}[\langle b|r^{i}(0)|d\rangle\langle d|r^{j}(0)|b\rangle e^{i\omega_{bd}(\tau-\tau')}
-\langle b|r^{j}(0)|d\rangle\langle d|r^{i}(0)|b\rangle e^{-i\omega_{bd}(\tau-\tau')}]\;\nn\label{chiA} \\
\eea
where $\omega_{bd}=\omega_{b}-\omega_{d}$ and the sum extends over a complete set of atomic states.

\section{Rate of change of the energy of a static atom}

Assume that an atom is placed static in a thermal bath with temperature $T$ in the cosmic string spacetime. In the cylindrical coordinates we use,
the position of the atom is denoted by $x(\tau)=(t(\tau),\rho,\theta,\phi)$ where $\rho,\theta,\phi$ are constants. As we have shown in the preceding  Section,
in order to calculate the average rate of change of the atomic energy, details on the two statistical functions of the field are indispensable.
Combine Eqs.~(\ref{relation 12}), (\ref{relation 34}) with Eq.~(\ref{Axi}), and then we get
\bea
A_{\rho}(t,\vec{x})&=&\frac{1}{\sqrt{2}}\int d \mu_j[(c_{+j}f_{+j}+c_{-j}f_{-j})+(c_{+j}^{\dag}f_{+j}^{\ast}+c_{-j}^{\dag}f_{-j}^{\ast})]\;,
\label{Arho}\\
A_{\theta}(t,\vec{x})&=&-\frac{i\rho}{\sqrt{2}}\int d \mu_j[(c_{+j}f_{+j}-c_{-j}f_{-j})-(c_{+j}^{\dag}f_{+j}^{\ast}-c_{-j}^{\dag}f_{-j}^{\ast})]\;,\\
A_{z,t}(t,\vec{x})&=&\int d \mu_j[c_{3j,0j}f_{0j}+c_{3j,0j}^{\dag}f_{0j}^{\ast}]\;,
\eea
where we have used the abbreviations $c_{+j}(t)\leftrightarrow c_{+j}$ and $f_{\xi j}(\vec{x})\leftrightarrow f_{\xi j}$. Making  use of
the relation $E_i=A_{0;i}-A_{i;0}$ leads to
\beq
\langle\beta|E_{i}(x)E_{j}(x')|\beta\rangle=
\partial_{0}\partial_{0}'\langle\beta|A_{i}(x)A_{j}(x')|\beta\rangle
+\partial_{i}\partial_{j}'\langle\beta|A_{0}(x)A_{0}(x')|\beta\rangle\;.
\eeq
The average value of an arbitrary operator, $G$, over the thermal state $|\beta\rangle$, can be obtained by using the following formula
\beq
\langle\beta|G|\beta\rangle=\frac{tr[\rho G]}{tr[\rho]}\;,\label{formula for thermal average G }
\eeq
where $\rho=e^{-\beta H_{F}}$ with $\beta=T^{-1}$ being the density matrix. Combining Eqs.~(\ref{Arho})-(\ref{formula for thermal average G }) with
Eq.~(\ref{definition for CF}), the non-zero components of the correlation functions of the field are found to be
\bea
C_{11}^{F\beta}(x(\tau),x(\tau'))&=&\frac{\nu}{8\pi^{2}}\int d \mu_j
\coth(\omega/T)\cos(\mathcal{}\omega(t-t'))\nn\\&&\quad\times
\biggl[\frac{\omega}{2}(J_{|\nu m+1|}^{2}(k_{\perp}\rho)+J_{|\nu m-1|}^{2}(k_{\perp}\rho))
-\frac{1}{\omega}\biggl(\frac{d J_{|\nu m|}(k_{\perp}\rho)}{d\rho}\biggr)^2\biggr]\;,\label{cf11}\\
C_{22}^{F\beta}(x(\tau),x(\tau'))&=&\frac{\nu\rho^2}{8\pi^{2}}\int d \mu_j
\coth(\omega/T)\cos(\mathcal{}\omega(t-t'))\nn\\&&\quad\times
\biggl[\frac{\omega}{2}(J_{|\nu m+1|}^{2}(k_{\perp}\rho)+J_{|\nu m-1|}^{2}(k_{\perp}\rho))
-\frac{1}{\omega}\frac{\nu^2m^2}{\rho^2}J^2_{|\nu m|}(k_{\perp}\rho)\biggr]\;,\\
C_{33}^{F\beta}(x(\tau),x(\tau'))&=&\frac{\nu}{8\pi^{2}}\int d \mu_j\;\frac{k_{\perp}^2}{\omega}\coth(\omega/T)J_{|\nu m|}^{2}(k_{\perp}\rho)
\cos(\mathcal{}\omega(t-t'))\;.\label{cf33}
\eea
Similarly, a combination of Eqs.~(\ref{Arho})-(\ref{formula for thermal average G }) with
Eq.~(\ref{definition for ChiF}) gives the non-zero components of the susceptibility functions of the field
\bea
\chi_{11}^{F\beta}(x(\tau),x(\tau'))&=&-\frac{i\nu}{8\pi^{2}}\int d \mu_j\sin(\mathcal{}\omega(t-t'))\nn\\&&\quad\times
\biggl[\frac{\omega}{2}(J_{|\nu m+1|}^{2}(k_{\perp}\rho)+J_{|\nu m-1|}^{2}(k_{\perp}\rho))
-\frac{1}{\omega}\biggl(\frac{d J_{|\nu m|}(k_{\perp}\rho)}{d\rho}\biggr)^2\biggr]\;,\label{chif11}\\
\chi_{22}^{F\beta}(x(\tau),x(\tau'))&=&-\frac{i\nu\rho^2}{8\pi^{2}}\int d \mu_j
\sin(\mathcal{}\omega(t-t'))\nn\\&&\quad\times
\biggl[\frac{\omega}{2}(J_{|\nu m+1|}^{2}(k_{\perp}\rho)+J_{|\nu m-1|}^{2}(k_{\perp}\rho))
-\frac{1}{\omega}\frac{\nu^2m^2}{\rho^2}J^2_{|\nu m|}(k_{\perp}\rho)\biggr]\;,\\
\chi_{33}^{F\beta}(x(\tau),x(\tau'))&=&-\frac{i\nu}{8\pi^{2}}\int d \mu_j \frac{k_{\perp}^2}{\omega}J_{|\nu m|}^{2}(k_{\perp}\rho)
\sin(\mathcal{}\omega(t-t'))\;.\label{chif33}
\eea

Insert the correlation functions of the field (Eqs.~(\ref{cf11})-(\ref{cf33})) and the antisymmetric statistical functions
of the atom (Eq.~(\ref{chiA})) into Eq.~(\ref{formula for tf contribution}), assume that $\tau-\tau_0\rightarrow\infty$,
make the coordinate transformation, $k_{\perp}=\omega \sin\alpha, k_3=\omega \cos\alpha$ in which $\alpha\in[0,\pi]$,
$\omega\in[0,\infty)$, and  then we obtain, after some lengthy simplifications, the contributions of thermal fluctuations to the average
rate of change of the atomic energy
%\bea
%\biggr\langle \frac{dH_{A}(\tau)}{d\tau}\biggr\rangle_{tf}&=&\frac{e^{2}\nu}{2\pi^{2}}\sum_{\omega_{bd}<0}\omega^4_{bd}
%[|r_\rho|_{bd}^2f_-(|\omega_{bd}|,\alpha)+|r_\theta|_{bd}^2f_+(|\omega_{bd}|,\alpha)+|r_z|_{bd}^2g(|\omega_{bd}|,\alpha)]\nn\\&&\quad\quad\quad
%\biggl(\frac{1}{2}+\frac{1}{e^{\frac{|\omega_{bd}|}{T}}-1}\biggr)\nn\\&
%-&\frac{\nu e^{2}}{2\pi^{2}}\sum_{\omega_{bd}>0}\omega^4_{bd}
%[|r_\rho|_{bd}^2f_-(\omega_{bd},\alpha)+|r_\theta|_{bd}^2f_+(\omega_{bd},\alpha)+|r_z|_{bd}^2g(\omega_{bd},\alpha)]
%\biggl(\frac{1}{2}+\frac{1}{e^{\frac{\omega_{bd}}{T}}-1}\biggr)\;.
%\eea
\bea
\biggr\langle \frac{dH_{A}(\tau)}{d\tau}\biggr\rangle_{tf}&=&-\frac{e^{2}}{3\pi}\sum_{\omega_{bd}>0}\omega^4_{bd}
|\langle b|\mathrm{r}_i(0)|d\rangle|^2f_i(\omega_{bd},\rho,\nu)\biggl(\frac{1}{2}+\frac{1}{e^{\omega_{bd}/T}-1}\biggr)\nn\\&&
+\frac{e^{2}}{3\pi}\sum_{\omega_{bd}<0}\omega^4_{bd}
|\langle b|\mathrm{r}_i(0)|d\rangle|^2f_i(|\omega_{bd}|,\rho,\nu)\biggl(\frac{1}{2}+\frac{1}{e^{|\omega_{bd}|/T}-1}\biggr)\;,
\label{contribution-tf}
\eea
%where $\mathrm{r}_i$ ($i=1,2,3$) denotes the interval of the components of the dipole's displacement vector, and
where we have defined
\bea
f_1(\omega,\rho,\nu)&=&\frac{3\nu}{4}\sum_m\int_0^{1}dt\frac{t}{\sqrt{1-t^2}}[(2-t^2)J^2_{|\nu m+1|}(\omega\rho t)
+t^2J_{|\nu m|+1}(\omega\rho t)J_{|\nu m|-1}(\omega\rho t)]\;,\label{definition f1}\nn\\ \\
f_2(\omega,\rho,\nu)&=&\frac{3\nu}{4}\sum_m\int_0^{1}dt\frac{t}{\sqrt{1-t^2}}[(2-t^2)J^2_{|\nu m+1|}(\omega\rho t)
-t^2J_{|\nu m|+1}(\omega\rho t)J_{|\nu m|-1}(\omega\rho t)]\;,\label{definition f2}\nn\\ \\
f_3(\omega,\rho,\nu)&=&\frac{3\nu}{2}\sum_m\int_0^{1}dt\frac{t^3}{\sqrt{1-t^2}}J^2_{|\nu m|}(\omega\rho t)\;.\label{definition f3}
\eea
In obtaining the above results, we have used the following properties of the BesselJ functions:
\bea
&&\sum_m J^2_{|\nu m+1|}(x)=\sum_m J^2_{|\nu m-1|}(x)\;,\\
&&\sum_m J^2_{|\nu m|+1}(x)+\sum_m J^2_{|\nu m|-1}(x)=2\sum_m J^2_{|\nu m+1|}(x)\;, (\nu\geq1)%\\
%&&J_{\lambda-1}(x)+J_{\lambda+1}(x)=\frac{2\lambda}{x}J_{\lambda}(x)\;,\\
%&&J_{\lambda-1}(x)-J_{\lambda+1}(x)=2\frac{d J_{\lambda}(x)}{dx}\;.
\eea
It is easy to show that functions $f_i(\omega,\rho,\nu)$ are always positive. For an atom
in the excited state, only the first term in Eq.~(\ref{contribution-tf}), which is negative, contributes, while for an atom
 in the ground state, only the second term in Eq.~(\ref{contribution-tf}), which is positive, contributes, i.e., the
thermal fluctuations always de-excite an atom in the excited state and excite it in the ground state. This is
similar to what happens to an atom  in Minkowski spacetime with no boundaries~\cite{Audretsch94}.
However, there are also some sharp differences between the two cases. Obviously, as can be seen from Eq.~(\ref{contribution-tf}),
in the cosmic string spacetime, the contribution of thermal fluctuations depends on the polarization and the position of the atom,
which is similar to  a static atom  in the Minkowski spacetime with boundaries~\cite{Yu-na061,Yu-na062,Yu-th0912},
while in a free Minkowski spactime with no boundaries, the contribution of thermal fluctuations  does not depend on the polarization and
position of the atom~\cite{Audretsch94}.

Similarly, plug the correlation functions of the field (Eqs.~(\ref{chif11})-(\ref{chif33})) and the symmetric statistical
function (Eq.~(\ref{cA})) of the atom into Eq.~(\ref{formula for rr contribution}), do some simplifications, and then we obtain
the contribution of radiation reaction to the average rate of change of the atomic energy,
\bea
\biggr\langle \frac{dH_{A}(\tau)}{d\tau}\biggr\rangle_{rr}&=&-\frac{e^{2}}{6\pi}
\sum_{\omega_{bd}>0}\omega^4_{bd}|\langle b|\mathrm{r}_i(0)|d\rangle|^2f_i(\omega_{bd},\rho,\nu)\nn\\&&
-\frac{e^{2}}{6\pi}\sum_{\omega_{bd}<0}\omega^4_{bd}
|\langle b|\mathrm{r}_i(0)|d\rangle|^2f_i(|\omega_{bd}|,\rho,\nu)
\;.
\label{contribution-rr}
\eea
For both the ground and the excited-state atoms, the contribution of  the radiation reaction is always negative. So just as in
a free Minkowski spacetime~\cite{Audretsch94}, radiation reaction always diminishes the atomic energy. Comparing this
result with the contribution of thermal fluctuations, Eq.~(\ref{contribution-tf}), we find that both contributions of
thermal fluctuations and radiation reaction depend on the polarization and position of the atom.

Adding up Eqs.~(\ref{contribution-tf}) and (\ref{contribution-rr}), we arrive at the total average rate of change of the atomic energy,
\bea
\biggr\langle \frac{dH_{A}(\tau)}{d\tau}\biggr\rangle_{tot}&=&-\frac{e^{2}}{3\pi}\sum_{\omega_{bd}>0}\omega^4_{bd}
|\langle b|\mathrm{r}_i(0)|d\rangle|^2f_i(\omega_{bd},\rho,\nu)\biggl(1+\frac{1}{e^{\omega_{bd}/T}-1}\biggr)\nn\\&&
+\frac{e^{2}}{3\pi}\sum_{\omega_{bd}<0}\omega^4_{bd}
|\langle b|\mathrm{r}_i(0)|d\rangle|^2f_i(|\omega_{bd}|,\rho,\nu)\frac{1}{e^{|\omega_{bd}|/T}-1}
\;.
\label{total rate}
\eea
For an atom  in the excited state, the first term, which is negative, contributes. It describes the
spontaneous emission rate of the excited atom immersed in a thermal bath in the cosmic string spacetime. For
an atom  in the ground state, the second term contributes and it is always positive. It describes
the spontaneous excitation rate of the atom. This  is clearly distinct from the transition rate of an inertial atom
 in the ground state in vacuum,
\bea
\biggr\langle \frac{dH_{A}(\tau)}{d\tau}\biggr\rangle_{tot}^{vac}=
-\frac{e^{2}}{3\pi}\sum_{\omega_{bd}>0}\omega^4_{bd}
|\langle b|\mathrm{r}_i(0)|d\rangle|^2f_i(\omega_{bd},\rho,\nu)\;,
\label{total rate-vaccum}
\eea
which is obtained by taking $T=0$ in Eq.~(\ref{total rate}). Obviously, the rate of change of the ground state
atom reduces to zero as a result of the complete cancelation of the contributions of vacuum fluctuations and radiation
reaction, i.e., for a ground-state atom placed in a vacuum in the cosmic string spacetime, no spontaneous excitation
occurs.

Generally,  analytical expressions for the functions $f_i(\omega,\rho,\nu)$ are not easy to find, but in some special cases, approximate analytical results are obtainable. We will examine these cases in the following.

\subsection{ The case for $\nu=1$.}
The case when $\nu=1$ corresponds to a flat spacetime without cosmic strings.  As a result of the following properties of the BesselJ function,
\beq
\sum_m J^2_{|m|}(x)=1\;,\quad\quad \sum_m J_{|m|+1}(x)J_{|m|-1}(x)=0\;,
\eeq
$f_i(\omega,\rho,\nu)=1$ ($i=1,2,3$). So, the contributions of thermal fluctuations and radiation reaction to the
average rate of change of the atomic energy reduce to
\bea
\biggr\langle \frac{dH_{A}(\tau)}{d\tau}\biggr\rangle_{tf}&=&-\frac{e^{2}}{3\pi}\sum_{\omega_{bd}>0}\omega^4_{bd}
|\langle b|\textbf{r}(0)|d\rangle|^2\biggl(\frac{1}{2}+\frac{1}{e^{\omega_{bd}/T}-1}\biggr)\nn\\&&
+\frac{e^{2}}{3\pi}\sum_{\omega_{bd}<0}\omega^4_{bd}
|\langle b|\textbf{r}(0)|d\rangle|^2\biggl(\frac{1}{2}+\frac{1}{e^{|\omega_{bd}|/T}-1}\biggr)
\;,
\label{Minkowski-contribution-tf}
\eea
\bea
\biggr\langle \frac{dH_{A}(\tau)}{d\tau}\biggr\rangle_{rr}&=&-\frac{e^{2}}{3\pi}\sum_{\omega_{bd}>0}\omega^4_{bd}
|\langle b|\textbf{r}(0)|d\rangle|^2\biggl(\frac{1}{2}+\frac{1}{e^{\omega_{bd}/T}-1}\biggr)\nn\\&&
-\frac{e^2}{3\pi}\sum_{\omega_{bd}<0}\omega^4_{bd}
|\langle b|\textbf{r}(0)|d\rangle|^2\biggl(\frac{1}{2}+\frac{1}{e^{|\omega_{bd}|/T}-1}\biggr)\;,
\label{Minkowski-contribution-rr}
\eea
where we have used the abbreviation,
\beq
|\langle b|\textbf{r}(0)|d\rangle|^2=\sum_i|\langle b|\mathrm{r}_i(0)|d\rangle|^2\;.
\eeq
Thus the total rate of change of the atomic energy becomes
\bea
\biggr\langle \frac{d H_{A}(\tau)}{d\tau}\biggr\rangle_{tot} &=&-\frac{e^2}{3\pi}\sum_{\omega_{bd}>0}\omega^4_{bd}
|\langle b|\textbf{r}(0)|d\rangle|^2\biggl(1+\frac{1}{e^{\omega_{bd}/T}-1}\biggr)\nn\\&&
+\frac{e^2}{3\pi}\sum_{\omega_{bd}<0}\omega^4_{bd}
|\langle b|\textbf{r}(0)|d\rangle|^2\frac{1}{e^{|\omega_{bd}|/T}-1}
\label{Minkowski-contribution-total}
\eea
which is just the average rate of change of an inertial atom placed in a thermal bath with temperature $T$ in a free Minkowski spacetime,
i.e., when $\nu=1$, the result in Minkowski spacetime is recovered as expected.

\subsection{The case for $\nu>1$. }

Let us note that when $\omega\rho\ll1$, one has
\beq
f_1(\omega,\rho,\nu)\approx f_2(\omega,\rho,\nu)\approx\frac{3\nu^2(\nu+1)}{\Gamma[2\nu+2]}(\omega\rho)^{2(\nu-1)}\equiv g(\omega\rho,\nu)\;,
\quad f_3(\omega,\rho,\nu)\approx\nu\;.
\eeq
So, when $\rho\ll\omega_{max}^{-1}$ where $\omega_{max}$ denotes the largest energy gap between two levels of the atom,  the contribution of thermal fluctuations reduces to
\bea
\biggr\langle \frac{dH_{A}(\tau)}{d\tau}\biggr\rangle_{tf}&\approx&-\frac{e^{2}}{3\pi}\sum_{\omega_{bd}>0}\omega^{4}_{bd}[
|\langle b|\mathrm{r}_{\perp}(0)|d\rangle|^2 g(\omega_{bd}\rho,\nu)+|\langle b|\mathrm{r}_{z}(0)|d\rangle|^2\nu]
\biggl(\frac{1}{2}+\frac{1}{e^{\omega_{bd}/T}-1}\biggr)\nn\\&&
+\frac{e^{2}}{3\pi}\sum_{\omega_{bd}<0}\omega_{bd}^{4}[
|\langle b|\mathrm{r}_{\perp}(0)|d\rangle|^2 g(|\omega_{bd}|\rho,\nu)+
|\langle b|\mathrm{r}_z(0)|d\rangle|^2\nu]
\biggl(\frac{1}{2}+\frac{1}{e^{|\omega_{bd}|/T}-1}\biggr)\nn\\
\label{contribution-tf-near}
\eea
where we have defined
\beq
|\langle b|\mathrm{r}_{\perp}(0)|d\rangle|^2=\sum_{i=1}^2|\langle b|\mathrm{r}_{i}(0)|d\rangle|^2\;,
\eeq
and we call this region ($\rho\ll\omega_{max}^{-1}$) the near zone. The contribution of radiation reaction becomes
\bea
\biggr\langle \frac{dH_{A}(\tau)}{d\tau}\biggr\rangle_{rr}&\approx&-\frac{e^{2}}{6\pi}\sum_{\omega_{bd}>0}\omega^4_{bd}[
|\langle b|\mathrm{r}_{\perp}(0)|d\rangle|^2 g(\omega_{bd}\rho,\nu)+|\langle b|\mathrm{r}_{z}(0)|d\rangle|^2\nu]\nn\\&&
+\frac{e^{2}}{6\pi}\sum_{\omega_{bd}<0}\omega_{bd}^4[
|\langle b|\mathrm{r}_{\perp}(0)|d\rangle|^2 g(|\omega_{bd}|\rho,\nu)+|\langle b|\mathrm{r}_z(0)|d\rangle|^2\nu]\;.
\label{contribution-rr-near}
\eea
As a result, the total average rate of change of the atomic energy can be written as
\bea
\biggr\langle \frac{dH_{A}(\tau)}{d\tau}\biggr\rangle_{tot}&\approx&-\frac{e^{2}}{3\pi}\sum_{\omega_{bd}>0}\omega^4_{bd}[
|\langle b|\mathrm{r}_{\perp}(0)|d\rangle|^2 g(\omega_{bd}\rho,\nu)+|\langle b|\mathrm{r}_{z}(0)|d\rangle|^2\nu]
\biggl(1+\frac{1}{e^{\omega_{bd}/T}-1}\biggr)\nn\\&&
+\frac{e^{2}}{3\pi}\sum_{\omega_{bd}<0}\omega_{bd}^4[
|\langle b|\mathrm{r}_{\perp}(0)|d\rangle|^2 g(|\omega_{bd}|\rho,\nu)+
|\langle b|\mathrm{r}_z(0)|d\rangle|^2\nu]\frac{1}{e^{|\omega_{bd}|/T}-1}\;.
\label{contribution-tot-near}
\eea
This shows that when the atom is located in the near zone, the spontaneous emission rate
of the atom in the excited state and spontaneous excitation rate of that in the ground state are proportional to $(|\omega _{bd}|\rho )^{2(\nu -1)}\ll 1$.
As a result, the average rate of change of the energy of an atom polarizable perpendicular to the string is much smaller
than that in a free Minkowski spacetime, while for an atom polarizable parallel to the string, this rate is always slightly larger  as $\nu$ is slightly larger than $1$ for a GUT (grand unified theory) string. In other words, the deficit in angle in the cosmic string
spacetime slightly amplifies this rate.

When $\rho=0$, i.e., the atom is exactly located on the string,
%. Notice that $\sum_{m}J_{|\nu m|}^{2}(0)=J_{0}^{2}(0)=1$,
%and generally $\nu$ is a non-integer. So,
%\beq
%\sum_{m}J_{|\nu m+1|}^{2}(0)=\sum_{m}J_{|\nu m|-1}(0)J_{|\nu m|+1}(0)=0\;,
%\eeq
%which leads to
\beq
f_1(\omega,\rho,\nu)=f_2(\omega,\rho,\nu)=0\;,\quad\quad f_3(\omega,\rho,\nu)=\nu\;.
\eeq
Then the contributions of vacuum fluctuations and radiation reaction reduce to
\bea
\biggr\langle \frac{dH_{A}(\tau)}{d\tau}\biggr\rangle_{tf}&=&-\frac{\nu e^{2}}{3\pi}\sum_{\omega_{bd}>0}\omega^4_{bd}
|\langle b|\mathrm{r}_z(0)|d\rangle|^2\biggl(\frac{1}{2}+\frac{1}{e^{\omega_{bd}/T}-1}\biggr)\nn\\&&
+\frac{\nu e^{2}}{3\pi}\sum_{\omega_{bd}<0}\omega^4_{bd}
|\langle b|\mathrm{r}_z(0)|d\rangle|^2\biggl(\frac{1}{2}+\frac{1}{e^{|\omega_{bd}|/T}-1}\biggr)
\;,
\label{cs-rho-0-contribution-tf}
\eea
\bea
\biggr\langle \frac{dH_{A}(\tau)}{d\tau}\biggr\rangle_{rr}&=&-\frac{\nu e^{2}}{3\pi}\sum_{\omega_{bd}>0}\omega^4_{bd}
|\langle b|\mathrm{r}_z(0)|d\rangle|^2\biggl(\frac{1}{2}+\frac{1}{e^{\omega_{bd}/T}-1}\biggr)\nn\\&&
-\frac{\nu e^2}{3\pi}\sum_{\omega_{bd}<0}\omega^4_{bd}
|\langle b|\mathrm{r}_z(0)|d\rangle|^2\biggl(\frac{1}{2}+\frac{1}{e^{|\omega_{bd}|/T}-1}\biggr)\;.
\label{cs-rho-0-contribution-rr}
\eea
The above two equations show that thermal fluctuations and radiation reaction affect only atoms polarizable parallel
to the string and they have no effect on atoms polarizable perpendicular to the string. This can be traced back to the fact
that on the string, only the $z-$component of the electric field is nonzero. It is reminiscent of a perfect conducting boundary where
only component of the electric field which is perpendicular to the surface is non-zero. In this sense, the effect of a cosmic
string is very similar to that of a perfect conducting boundary. This is understandable since the cosmic string only modifies the global spacetime topology while leaving the local space flatness intact, which is pretty much the same as what a conducting boundary does to a flat space.
%This property is in sharp contrast to that in other spacetime\cite{}.

Adding up the above two equations, we obtain the total rate of change of the atomic energy,
\bea
\biggr\langle \frac{dH_{A}(\tau)}{d\tau}\biggr\rangle_{tot}&=&
-\frac{\nu e^{2}}{3\pi}\sum_{\omega_{bd}>0}\omega_{bd}^{4}|\langle b|\mathrm{r}_{z}(0)|d\rangle|^{2}\biggl(1+\frac{1}{e^{\omega_{bd}/T}-1}\biggr)\nn\\&&
-\frac{\nu e^{2}}{3\pi}\sum_{\omega_{bd}<0}\omega_{bd}^{4}|\langle b|\mathrm{r}_{z}(0)|d\rangle|^{2}\frac{1}{e^{|\omega_{bd}|/T}-1}\;.
\eea
This  shows that when the atom is located on the string, the average rate of change of the atomic energy depends crucially
on the polarization of the atom. For an atom in the excited state, spontaneous emission can occur only if it is polarizable
parallel to the string, whereas those which are only polarizable perpendicular to the string will remain in the excited states
and thus are stable. Meanwhile, the ground-state atoms can make a transition to excited states only if they are polarizable parallel
to the string. Even if  immersed in a thermal bath, ground state atoms  polarizable only perpendicular to the string are stable
as if they were in a vacuum.  This is in sharp contrast to the case of a thermal bath in the Minkowski spacetime, where spontaneous
emission takes place for excited atoms polarizable in any direction, and spontaneous excitation  occurs for any polarizable ground
state atoms (see Eq.~(\ref{Minkowski-contribution-total})).
It is interesting to note that similar properties also appear in the case of an atom located near
a perfect conducting plate in Minkowski spacetime, in which the rate of change of the energy of an atom polarizable parallel to
surface of the plate vanishes when the atom-surface distance approaches zero, while the rate for an atom polarizable perpendicular
to the surface of the conducting plate doesn't vanish~\cite{Yu-na062}. This suggests that effect of a deficit angle induced by a cosmic string is similar to that of a reflecting boundary in a flat spacetime. This is reasonable from a physical point of view since the cosmic string spacetime is locally flat and what distinguishes it from a Minkowski spacetime is its nontrivial topology characterized by the deficit angle.

When $\omega\rho\gg1$,  we first do the $t$-integrals in Eqs.~(\ref{definition f1})-(\ref{definition f3}),
and then in the limit $\omega\rho\gg1$ we can cut off the infinite $m-$summation by $|m|\leq\omega\rho\nu^{-1}$, which  results in
\beq
f_i(\omega,\rho,\nu)\approx1+\frac{3\nu}{4\omega\rho},\;(i=1,3), \quad\; f_2(\omega,\rho,\nu)\approx1-\frac{\nu^2}{4\omega^2\rho^2}\;.
\label{far zone fi}
\eeq
%\beq
%f_i(\omega,\rho,\nu)\sim1,\;(i=1,2,3).
%\eeq
As a result, for an atom located in the region, $\rho\gg\omega_{min}^{-1}$, where $\omega_{min}$ denotes the smallest energy gap between two levels of the atom, the contributions of thermal fluctuations and radiation reaction to the average rate of change
of the atomic energy reduce to
\bea
\biggr\langle \frac{dH_{A}(\tau)}{d\tau}\biggr\rangle_{tf}&\approx&-\frac{e^{2}}{3\pi}\sum_{\omega_{bd}>0}\omega^4_{bd}
|\langle b|\textbf{r}(0)|d\rangle|^2\biggl(\frac{1}{2}+\frac{1}{e^{\omega_{bd}/T}-1}\biggr)\nn\\&&
+\frac{e^{2}}{3\pi}\sum_{\omega_{bd}<0}\omega^4_{bd}
|\langle b|\textbf{r}(0)|d\rangle|^2\biggl(\frac{1}{2}+\frac{1}{e^{|\omega_{bd}|/T}-1}\biggr)
\;,
\label{far zone-vf}
\eea
\bea
\biggr\langle \frac{dH_{A}(\tau)}{d\tau}\biggr\rangle_{rr}&\approx&-\frac{e^{2}}{3\pi}\sum_{\omega_{bd}>0}\omega^4_{bd}
|\langle b|\textbf{r}(0)|d\rangle|^2\biggl(\frac{1}{2}+\frac{1}{e^{\omega_{bd}/T}-1}\biggr)\nn\\&&
-\frac{e^2}{3\pi}\sum_{\omega_{bd}<0}\omega^4_{bd}
|\langle b|\textbf{r}(0)|d\rangle|^2\biggl(\frac{1}{2}+\frac{1}{e^{|\omega_{bd}|/T}-1}\biggr)\;,
\label{far zone-rr}
\eea
and thus the total rate of change of the atomic energy becomes
\bea
\biggr\langle \frac{d H_{A}(\tau)}{d\tau}\biggr\rangle_{tot} &\approx&-\frac{e^2}{3\pi}\sum_{\omega_{bd}>0}\omega^4_{bd}
|\langle b|\textbf{r}(0)|d\rangle|^2\biggl(1+\frac{1}{e^{\omega_{bd}/T}-1}\biggr)\nn\\&&
+\frac{e^2}{3\pi}\sum_{\omega_{bd}<0}\omega^4_{bd}
|\langle b|\textbf{r}(0)|d\rangle|^2\frac{1}{e^{|\omega_{bd}|/T}-1}\;.
\eea
We call the region, $\rho\gg\omega_{min}^{-1}$, the far zone. In the above three equations, we have only kept the leading terms.
For an atom polarizable along the radial direction or parallel
to the $z-$direction, the rate is actually slightly larger than that in a Minkowski spacetime as a positive term proportional to
$\rho^{-1}$ exists going to the next order (see Eq.~(\ref{far zone fi})), and for an atom polarizable along the tangential
direction, the rate is slightly smaller than that in a Minkowski spacetime because $f_2(\omega,\rho,\nu)$ is actually amended by a
negative term proportional to $\rho^{-2}$ (see Eq.~(\ref{far zone fi})).
The above results show that in the far zone where the atom-string distance is much larger than the longest transition wavelength of the atom,
the average rate of change of the atomic energy approximates to that in a Minkowski spacetime. This  is similar to the
behavior of the rate of a static atom placed far away from a perfect reflecting boundary in Minkowski spacetime as the boundary
effect vanishes at infinity~\cite{Yu-na062}. This is in accordance with our observation that the deficit angle in the cosmic string spacetime affects the fields the atom couples to in a way which is very similar to a reflecting boundary in Minkowski spacetime. Compare this result with that of a static atom coupled to quantum scalar field
in the cosmic string spacetime~\cite{Audretsch952}, we find that the conclusions are consistent, as in the latter case, the
decay rate of a static atom coupled to quantum scalar field in the cosmic string spacetime also approaches the
result in a free Minkowski spacetime  at infinity.

It is worth pointing out here that the above approximations in the present case  do not hold  when $\nu=1$ which have already been  discussed in the preceding subsection (case A).  For a generic atom-string distance, an analytical analysis is impossible  for the average rate of change of the atomic energy. %and it generally depends on the parameter $\nu$ and the atom-string distance.
So, instead, we now give some  results of numerical  in this case. The following figures show how the rate
of change of the atomic energy varies  as a function of the parameter $\nu$ and the atom-string distance.  We consider the ratio $\frac{\Gamma_{cs}}{\Gamma_{0}}$ with $\Gamma_{cs}$ and $\Gamma_{0}$
denoting the average rates of change of energy of a two-level atom in the cosmic string spacetime and the Minkowski spacetime
respectively.  The spacing between the two levels of the atom is represented by $\omega_0$.
\begin{figure}
\centering
\subfigure[The case for an atom polarizable along the radial direction.]{\includegraphics[scale=0.75]{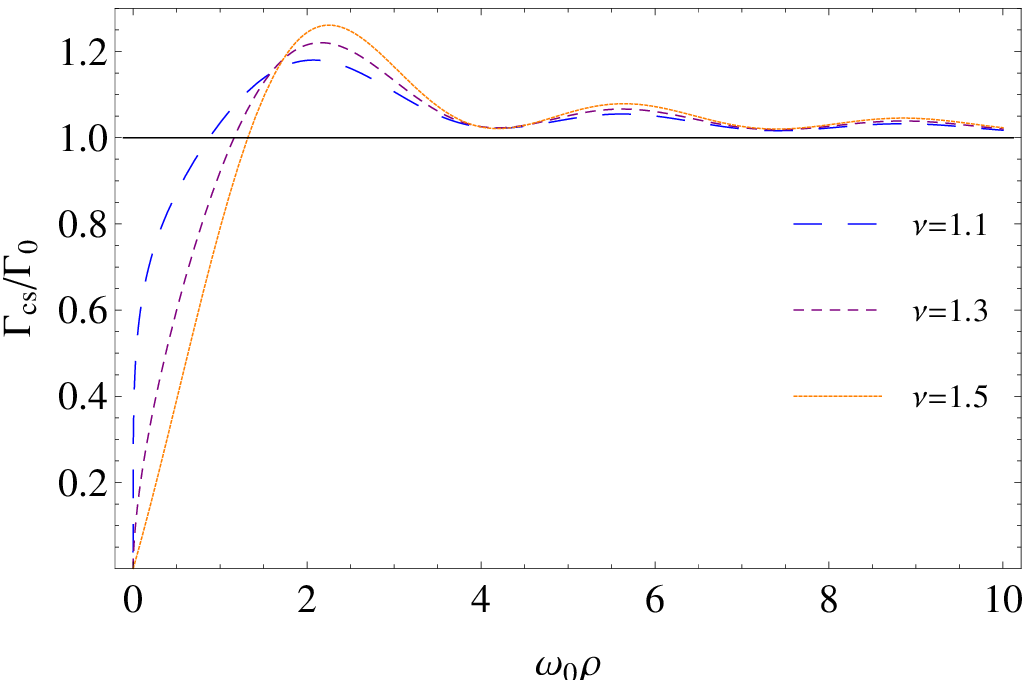}\label{figure1}}
\subfigure[The case for an atom polarizable along the tangential direction.]{\includegraphics[scale=0.75]{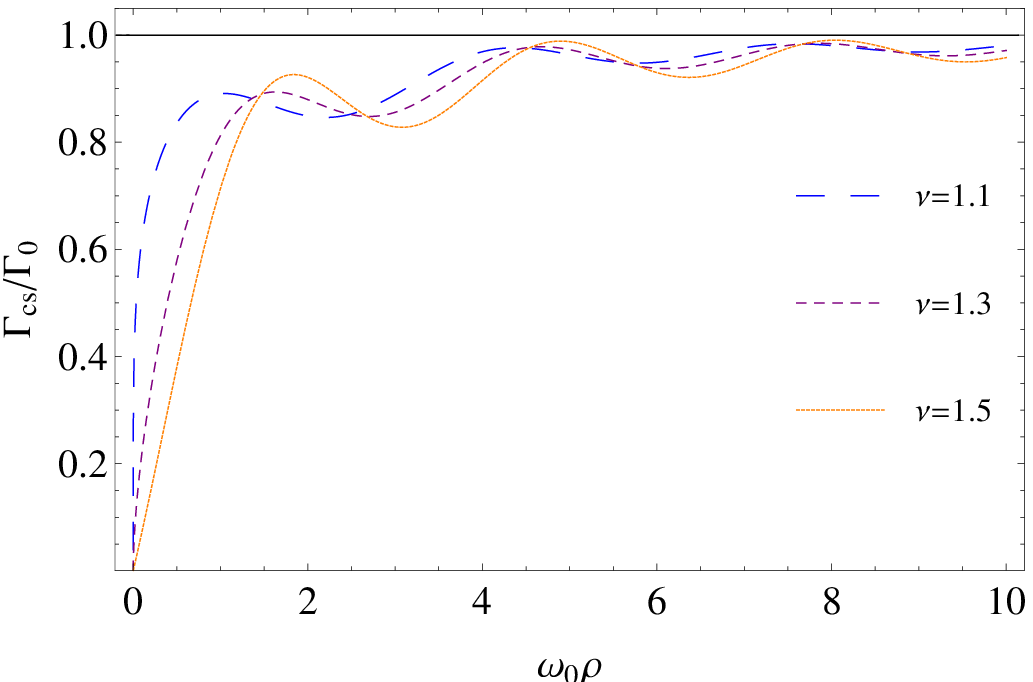}\label{figure2}}
\centering
\subfigure[The case for an atom polarizable parallel to the string.]{\includegraphics[scale=0.75]{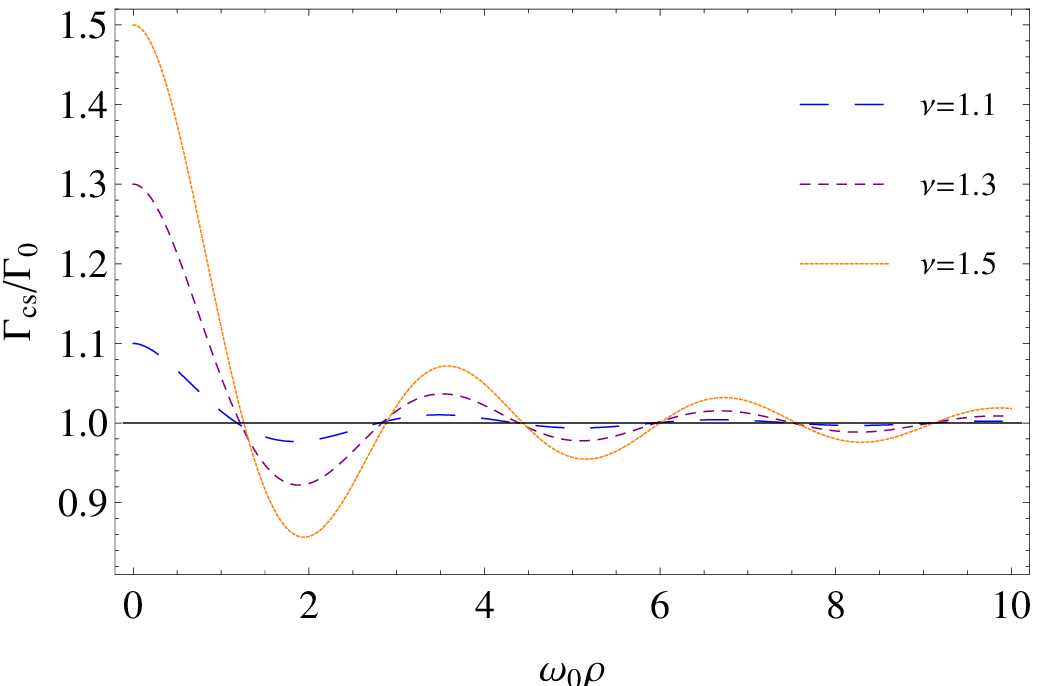}\label{figure3}}
\subfigure[The case for an atom polarizable isotropically.]{\includegraphics[scale=0.75]{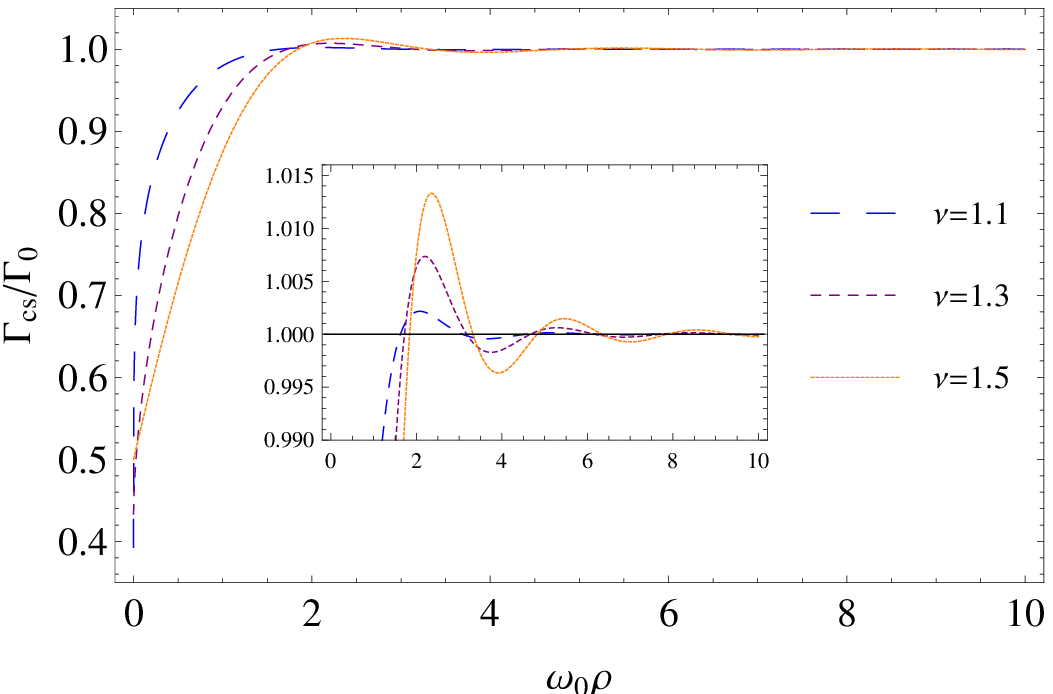}\label{figure4}}
\caption{Ratio between the rate of change of a two-level static atom in the cosmic string spacetime and that in a free Minkowski spacetime.}
\end{figure}

As shown in the four figures, the relative rate $\frac{\Gamma_{cs}}{\Gamma_{0}}$ for a static atom generally oscillates with
the atom-string distance, and the amplitude of oscillation decreases with increasing atom-string distance. Moreover, the oscillation
is more severe for larger $\nu$, i.e., larger deficit in the angle induce more severe oscillation. For a two-level atom
polarizable along the radial direction,  the rate of change of the atomic energy
in the cosmic string spacetime is smaller than that in Minkowski spacetime when the atom is located very close to the string, which means that the atomic energy varies slower than
in a free Minkowski spacetime. When the atom-string distance exceeds a critical value,  the average rate of change of energy in the cosmic string  overtakes that in a free Minkowski spacetime as indicated by that the relative rate now becomes larger than unity (see Figure.~\ref{figure1}), although the relative rate still oscillates with the distance.  The rate of change of the atomic energy approaches that in a Minkowski spacetime as the atom-string distance becomes larger and larger. For an atom polarizable in
the tangential direction, (see Figure.~\ref{figure2}), the rate of change of the atomic energy is always smaller than that in a free Minkowski
spacetime, and the difference becomes smaller with the increase of the atom-string distance. For an atom polarizable parallel to the string
(see Figure.~\ref{figure3}), the rate of change of the atomic energy can be larger or smaller than that in a free Minkowski spacetime as the
ratio $\frac{\Gamma_{cs}}{\Gamma_{0}}$ oscillates around unity as the atom-string distance varies. Notice that here the
numerical results are consistent with our previous analytical analysis on the average rate of change of the energy of an atom located very
close to the string in that for an atom polarizable perpendicular to the string,
the rate is proportional to $\rho^{2(\nu-1)}\sim0$, and for an atom polarizable parallel to the string, the rate is proportional to $\nu$.  We  show also the ratio $\frac{\Gamma_{cs}}{\Gamma_{0}}$ for an isotropically polarizable atom in Figure.~\ref{figure4}, and one can see that it also oscillates around unity,
but the amplitude of oscillation is much smaller than the ratio of an atom polarizable parallel to the string (see Figure.~\ref{figure3}).

%\textbf{Finally, we give some comments. As shown both analytically and numerically, the behavior of the average rate of the atomic
%energy in the cosmic string spacetime is similar to that in a Minkowski spacetime with boundaries. This dependence of the rate on
%the atom-string distance is induced by the special topology configuration of the comic string spacetime. Due to the deficit in angle,
%the topology is nontrivial, thus modes traveling in the cosmic string spacetime are restricted as opposed to in a free Minkowski spacetime,
%as a result, boundary effects appear.}

\section{CONCLUSIONS}

We have studied the average rate of change of a multilevel static atom coupled to quantum
electromagnetic field in a thermal bath in the cosmic string spacetime. We separately calculate the contributions of thermal fluctuations of the field and
radiation reaction of the atom to the average rate of change of the atomic energy. We analyze  the behavior of the transition rates analytically in both the near zone and the far zone and numerically for a generic atom-string distance.
We find that the transition rates are  crucially dependent on  the atom-string distance and
 polarization of the atom and they in general oscillate as the atom-string distance varies. Moreover, the atomic transition rates in the cosmic string spacetime can be larger or smaller than those in Minkowski spacetime contingent upon the atomic polarization and position, meaning the transition rates can be either enhanced or weakened by the cosmic string.  In particular, when located on the string, ground-state atoms can transition to excited states only if they are polarizable parallel
to the string, whereas  ground state atoms  polarizable only perpendicular to the string are stable
as if they were in a vacuum, even if  they are immersed in a thermal bath.  This feature can be attributed to the fact
that on the string, only the $z-$component of the electric field is nonzero and it is reminiscent of a perfect conducting boundary where
only component of the electric field which is perpendicular to the surface is non-zero. In this sense, the effect of a cosmic
string is very similar to that of a perfect conducting boundary. This does not come as a surprise  since the cosmic string only modifies the global spacetime topology while leaving the local space flatness intact in a similar way as what a conducting boundary does to a flat space.

\addcontentsline{toc}{chapter}{Acknowledgment}
\begin{acknowledgments}
This work was supported in part by the NSFC under
Grants No. 11375092, No. 11435006, and No. 11405091;
the SRFDP under Grant No. 20124306110001; the Zhejiang
Provincial Natural Science Foundation of China under
Grant No. LQ14A050001; the Research Program of Ningbo
University under Grants No. E00829134702, No. xkzwl10,
and No. XYL14029; and the K. C. Wong Magna Fund in
Ningbo University.
\end{acknowledgments}


\baselineskip=16pt

\begin{thebibliography}{35}
\bibitem{Welton48} T. A. Welton, Phys. Rev. {\bf 74}, 1157 (1948).
\bibitem{Compagno83} G. Compagno, R. Passante and F. Persico, Phys. Lett. A {\bf 98}, 253 (1983).
\bibitem{Ackerhalt73} J. R. Ackerhalt, P. L. Knight and J. H. Eberly, Phys. Rev. Lett. {\bf 30}, 456 (1973).
\bibitem{Milonni88} P. W. Milonni, Phys. Scr. {\bf 21}, 102 (1988).
\bibitem{Milonni75} P. W. Milonni and W. A. Smith, Phys. Rev. A {\bf 11}, 814 (1975).
\bibitem{Vleck24} J. H. van Vleck, Phys. Rev. {\bf 24}, 330 (1924).
\bibitem{Dirac27} P. A. M. Dirac, Pro. Roy. Soc. Lond. A {\bf 114}, 243 (1927).
\bibitem{Senitzky73} I. R. Senitzky, Phys. Rev. Lett. 31, 955 (1973).
\bibitem{Milonni73} P. W. Milonni, J. R. Ackerhalt and W. A. Smith, Phys. Rev. Lett. {\bf 31}, 958 (1973).
\bibitem{Ackerhalt74} J. R. Ackerhalt and J. H. Eberly, Phys. Rev. D {\bf 10}, 3350 (1974).
\bibitem{Milonni752} P. W. Milonni, Phys. Rep. {\bf 25}, 1 (1975).
\bibitem{Dalibard8284} J. Dalibard, J. Dupont-Roc, and C. Cohen-Tannoudji, J. Phys. (Paris) {\bf 43}, 1617 (1982); ibid, {\bf 45}, 637 (1984).
\bibitem{Audretsch94} J. Audretsch and R. M\"{u}ller, Phys. Rev. A {\bf 50}, 1755 (1994); ibid, {\bf 52}, 629 (1995).
\bibitem{Audretsch951} J. Audretsch, R. M\"{u}ller, and M. Holzmann, Class. Quant. Grav. 12, 2927 (1995).
\bibitem{Passante98} R. Passante, Phys. Rev. A {\bf 57}, 1590 (1998).
\bibitem{Yu-na05} H. Yu and S. Lu, Phys. Rev. D {\bf 72}, 064022 (2005).
\bibitem{Yu-na061} Z. Zhu, H. Yu and S. Lu, Phys. Rev. D {\bf 73}, 107501 (2006).
\bibitem{Yu-na062} Z. Zhu and H. Yu, Phys. Rev. D {\bf 74}, 044032 (2006).
\bibitem{Yu-na07} Z. Zhu and H. Yu, Phys. Lett. B {\bf 645}, 459 (2007).
\bibitem{Rizzuto07} L. Rizzuto, Phys. Rev. A {\bf 76}, 062114 (2007).
\bibitem{Rizzuto09} L. Rizzuto and S. Spagnolo, Phys. Rev. A {\bf 79}, 062110 (2009);
                      ibid, J. Phys.: Conf. Ser. {\bf 161}, 012031 (2009).
\bibitem{Yu-na10} Z. Zhu and H. Yu, Phys. Rev. A {\bf 82}, 042108 (2010).
\bibitem{Rizzuto11} L. Rizzuto and S. Spagnolo, Phys. Scr., T {\bf 143}, 014021 (2011).
\bibitem{Yu-na12} W. Zhou and H. Yu, Phys. Rev. A {\bf 86}, 033841 (2012).
\bibitem{Tomazelli03} J. L. Tomazelli and L. C. Costa, Int. J. Mod. Phys. A {\bf 18}, 1079 (2003).
\bibitem{Yu-th0912} Z. Zhu and H. Yu, Phys. Rev. A {\bf 79}, 032902 (2009);
                    ibid, Phys. Rev. A {\bf 86}, 052508 (2012).
\bibitem{Yu-th10}W. She, H. Yu and Z. Zhu, Phys. Rev. A {\bf 81}, 012108 (2010).
\bibitem{Iliadakis} L. Iliadakis, U. Jasper, and J. Audretsch, Phys. Rev. D {\bf 51}, 2591 (1995).
\bibitem{Yu-cur07} W. Zhou and H. Yu, JHEP {\bf 4}, 024 (2007); H. Yu and W. Zhou, Phys. Rev. D {\bf 76}, 027503 (2007); ibid, {\bf 76}, 044023 (2007).
\bibitem{Yu-cur08} Z. Zhu and H. Yu, JHEP {\bf 2}, 033 (2008).
\bibitem{Yu-cur12} W. Zhou and H. Yu, Class. Quant. Grav. {\bf 29}, 085003 (2012); ibid, JHEP {\bf 10}, 172 (2012).
\bibitem{Vilenkin81} A. Vilenkin, Phys. Rev. D {\bf 23}, 852 (1981).
\bibitem{Aliev89} A. N. Aliev and D. V. Gal'tsov, Ann. Phys., NY {\bf 193}, 142 (1989).
\bibitem{Aliev93} A. N. Aliev, Class. Quant. Grav. {\bf 10}, 2531 (1993).
\bibitem{Saharian11} A. A. Saharian, A. S. Kotanjyan, Phys. Lett. B {\bf 713}, 133 (2012); ibid, Eur. Phys. J. C {\bf 71}, 1765 (2011).
\bibitem{Saharian121} E. R. Bezerra de Mello, A. A. Saharian, and A. Kh. Grigoryan, J. Phys. A: Math. Theor. {\bf 45}, 374011 (2012).
\bibitem{Saharian122} E. R. Bezerra de Mello, V. B. Bezerra, H. F. Mota and A. A. Saharian, Phys. Rev. D {\bf 86}, 065023 (2012).
\bibitem{Audretsch952} L. Iliadakis, U. Jasper, and J. Audretsch, Phys. Rev. D {\bf 51}, 2591 (1995).
\bibitem{Skarzhinsky94} V. D. Skarzhinsky, D. D. Harari, and U. Jasper, Phys. Rev. D {\bf 49}, 755 (1994).
\end{thebibliography}
\end{document}